# A Novel Route for the Inclusion of Metal Dopants in Silicon

Jules A. Gardener [1,*], Irving Liaw [2,**], Gabriel Aeppli [1], Ian W. Boyd [2], Richard J. Chater [3], Tim S. Jones [4], David S. McPhail [3], Gopinathan Sankar [5], A. Marshall Stoneham [1], Marcin Sikora [6,***], Geoff Thornton [7], and Sandrine Heutz [3, §]

[1] Department of Physics and London Centre for Nanotechnology, University College London, London WC1E 6BT, UK.

[2] Department of Electronic and Electrical Engineering and London Centre for Nanotechnology, University College London, London WC1E 6BT, UK.

[3] Department of Materials and London Centre for Nanotechnology, Imperial College London, London SW7 2AZ, UK.

[4] Department of Chemistry, University of Warwick, Coventry CV4 7AL, UK.

[5] Department of Chemistry and London Centre for Nanotechnology, University College London, London WC1E 6BT, UK.

[6] ESRF, 6 rue Jules Horowitz, BP 220, F-38043 Grenoble Cedex, France.

* Present address: Department of Physics, Harvard University, Cambridge, MA 02138, USA
** Present address: School of Chemistry, University of Melbourne, Victoria 3010, Australia.
*** Present address: Department of Solid State Physics, AGH University of Science and Technology, 30-059 Krakow, Poland.
[§] electronic mail: s.heutz@imperial.ac.uk

**We report a new method to introduce metal atoms into silicon wafers, using negligible thermal budget. Molecular thin films are irradiated with ultra-violet (UV) light releasing metal species into the semiconductor substrate. Secondary ion mass spectrometry (SIMS) and X-ray absorption spectroscopy (XAS) show that Mn is incorporated into Si as an interstitial dopant. We propose that our method can form the basis of a generic low-cost, low-temperature technology that could lead to the creation of ordered dopant arrays.**

Chemically versatile, minimal thermal budget approaches to the controlled doping of semiconductor wafers become increasingly important as feature sizes fall. Subjecting pre-patterned dopant regions to high temperatures enhances diffusion and is detrimental to interface definition. Ion implantation or metallic thin film precursors can be used to dope semiconductors, but these procedures rely on annealing to produce high quality wafers [1, 2]. Alternatively, growth methods such as molecular beam epitaxy (MBE) or metal organic chemical vapour deposition (MOCVD) can be utilised, although these are expensive and also require a high temperature step [3, 5]. Lateral control of dopant positioning can be achieved using scanning probe methods [6, 7] or, recently, annealing of chemisorbed organic films [8]. However, these approaches are limited by a low prospect of upscaling and restrictive dopant-substrate combinations respectively. Here, we demonstrate a new approach, using clean, cheap vacuum ultra-violet sources [9] to degrade metalorganic thin films on Si at temperatures close to room temperature. This releases the metal atoms which are subsequently incorporated into the substrate as dopants.

The molecules chosen are phthalocyanines (MPc, Figure 1 inset), which can form ordered films onto virtually any substrate [10]. Once deposited, these molecules remain intact, and so the ligands act to spatially separate the metal atoms. Important for the present application, MPcs can contain a vast range of species at their centre (most transition metals, rare earths, oxides, etc), and we envisage that the doping method demonstrated here for Mn in Si could be easily generalised to any dopant/substrate combination. Here, we focus on manganese phthalocyanine (MnPc) to make Si:Mn, a system with potential applications in spintronics [11, 12].

A B-doped Si(100) substrate (0.010 $\Omega$cm) was hydrogen passivated by immersing in HF (VWR, 50%) to remove the native oxide layer and immediately loaded into a Kurt J. Lesker vacuum deposition chamber (base pressure $8\times10^{-8}$ mbar). Manganese (II) phthalocyanine (Aldrich), which had been further purified by two cycles of gradient sublimation, was sublimed at 390-400 °C, producing α-phase crystalline films wherein the MnPc lie almost perpendicular to the substrate [13]. The final film thickness of 5 nm was monitored by an in-situ quartz crystal microbalance. The films were irradiated with 172 nm photons from a purpose-built array of four Xe excimer lamps, in a separate chamber backfilled to 5.0 mbar with nitrogen [9]. During irradiation (40 minutes, unless otherwise stated), an in-situ thermocouple showed the substrate remained at 25-60 °C. UV-Vis analyses were performed on irradiated MnPc films of glass substrates (to allow optical transmission) using a Perkin-Elmer Lambda 950 UV-Vis spectrometer. Secondary ion mass spectroscopy (SIMS) was performed on Si samples with an Atomika 6500 microprobe (1.25 keV, 30 nA Ar+ primary ion beam incident at 45° to the surface), whilst electron beam compensation was not required. X-ray absorption spectroscopy (XAS) data at the Mn K-edge were collected at the European Synchrotron Radiation Facility (Station ID26) in fluorescence mode using a 13 element Ge detector and 0.9 eV excitation energy resolution. After UV irradiation, a Piranha clean removed any Mn-rich surface residue prior to SIMS and XANES analyses, resulting in a uniform surface $SiO_2$ layer.

MPc's are efficient light absorbers due to $\pi$-$\pi$* transitions corresponding to excitation from the molecular ground state to the first and second excited states in the visible and near-UV regions respectively [14]. UV-Vis measurements allow us to monitor the rate at which the MnPc film is degraded during exposure to the UV lamp array, as shown in Figure 1. The spectrum of the initial MnPc film agrees with previous work [15], confirming its integrity before UV exposure. Upon increasing irradiation, the intensity of the characteristic MnPc absorption spectrum decreases in a controlled manner until after forty minutes it is no longer detectable, thus demonstrating that the molecules have been modified and/or removed. The disappearance of the absorption profile corresponds to photo-decomposition [16], as processing temperatures remain below 60 °C and so are insufficient to sublime the molecules or induce structural changes [17]. From energetic considerations, the first dissociation step is the rupturing of the lowest energy C-N bonds, resulting in the formation of pyrrole- and benzene-rich fragments, which themselves will be further decomposed by the 172 nm photons [18]. This destabilises the metal-ligand bond leading to the release of the metal atoms, whilst other UV-assisted mechanisms aid transfer to the substrate, as discussed below.

Further analysis of the chemical composition of the surface and bulk Si after forty minutes of UV exposure has been performed using SIMS. Initial studies showed a high Mn concentration accompanied with oxygen enrichment before the substrate was reached, followed by a decreasing manganese profile within the silicon. This is consistent with surface manganese oxide, in agreement with other reports [16] and the broad UV-Vis peak below 500 nm (Figure 1) [19]. Therefore, to avoid introduction of Mn due to beam-induced mixing [20], and to focus on the Mn that has been incorporated into Si exclusively by the UV process, all degraded samples were Pirhana cleaned to remove any Mn-rich surface contaminants. Figure 2.a shows a sample that has been UV-treated and subjected to a Piranha clean. Even after cleaning, we observe a stable Mn presence which extends for depths of up to 95 nm. Numerous analyses were performed on the bare wafer (for example, as shown in Figure 2.b) and on Pirhana cleaned as-deposited MnPc films, none of which showed any traces of Mn. This confirms that the UV (rather than chemical) process leads to Mn incorporation. We estimate a Mn concentration of $7 \times 10^{20}$ atoms/cm$^3$ from the SIMS profile in Figure 2.a using the relative sensitivity factors [21]. Spatial composition mapping highlights lateral inhomogeneities and Figure 2.a represents the maximum Mn concentration observed. The concentration of Mn atoms in a MnPc film can be approximated by assuming that the unit cell of MnPc is similar to that of α-phase CuPc, with dimensions a = 25.9 Å, b = 3.8 Å, c = 23.9 Å and β = 90.4° (with four metal atoms per unit cell) [22]. This concentration ($C_{MnPc}$) is calculated in Equation 1, where $V_{uc}$ is the volume of the unit cell (in cm$^3$).

$$C_{MnPc} = \frac{4}{V_{uc}} = \frac{4}{abc\sin\beta} = 1.7 \times 10^{21} \ atoms/cm^3 \quad [1]$$

The number of Mn atoms in a 1 cm x 1 cm surface region of a MnPc film of thickness $d_{film}$ (in cm, in this case 5 nm thick) is given by Equation 2.

$$N_{Mn} = C_{MnPc} d_{film} = 8.5 \times 10^{14} \ atoms/cm^2 \quad [2]$$

Deriving the implanted Mn concentration from over 30 analyses over a range of sample preparation runs and comparing to the initial Mn concentration calculated above yields an efficiency for doping of 65±15%.

The type of bonding of the Mn dopant in the Si lattice can be inferred from the value of the Mn diffusion coefficient, *D*. We estimate the diffusion coefficient using successive SIMS measurements (with time delays of one day) and the solution of Fick's equation, which gives the concentration *C* of a dopant as a function of depth, *x*, and time, *t*, after diffusion from a surface concentration *C*(0) as given in Equation 3.

$$C(x,t) = C(0)erfc\left(\frac{x}{2\sqrt{Dt}}\right) \quad [3]$$

We find $D \sim (0.8 \pm 0.1) \times 10^{-15}$ cm$^2$/s at room temperature, which is close to the reported value for interstitial Mn in Si as determined by deep level transient spectroscopy and the Hall effect (2.8x10$^{-15}$ cm$^2$/s) [23]. We note that the preferential formation of interstitial Mn is consistent with reports elsewhere [24, 25]. The relatively constant Mn concentration observed in Figure 2.a is contrary to the decrease of ~ 10% expected over a depth of 30 nm using equation 3. This is likely to be due to experimental limitation in the stability of the SIMS measurements and to the beam induced mixing of surface Mn into the wafer.

More detailed information about the local chemical environment of the dilute implantedMn can be obtained using XAS techniques. The total Mn K-fluorescence yield from a UV-treated MnPc film after cleaning and storing at 193 K, compared with bulk Mn, gives an Mn concentration of ~5 x $10^{20}$ atoms/$cm^3$, assuming similar X-ray penetration and escape depths, a value which is consistent with our SIMS data. The UV-treated MnPc film, fresh MnPc film, Mn foil and B20 cubic Mn silicide are compared in the X-ray absorption near edge spectroscopy (XANES) profiles shown in Figure 3.a. These post-normalisation absorption features show notable differences between the UV-treated and precursor films. A significant lowering of the X-ray absorption edge by ~8.5 eV with respect to the precursor film (represented by the dashed and dotted lines respectively) demonstrates that Mn is no longer bound to the organic ligand, consistent with our UV-Vis data (Figure 1). The formation of oxides, which typically show an absorption edge shift of 5-15 eV higher than bulk Mn [26] (region highlighted in grey) are also ruled out. Instead, the edge of the UV treated sample corresponds to Mn0, as it occurs at the same energy as metallic Mn. However, the higher energy X-ray absorption features differ markedly from the Mn foil, so the formation of Mn clusters is unlikely. Instead, the lineshape and edge position of the UV-treated sample strongly resemble the silicide reference, suggesting Mn atoms surrounded by Si. This is even more evident in the extended X-ray absorption fine structure (EXAFS) spectrum (Figure 3.b), where the oscillations seen for the silicide and our doped sample occur at similar positions, contrary to what is observed for the metal or MnPc film. Whilst our experimental results have established that the Mn atoms are in a neutral oxidation state surrounded by silicon, it is unlikely that they form pure, stable bulk silicides. This is seen in the subtle discrepancies in the XANES and EXAFS spectra, suggesting differences in the Mn-Si distance and/or co-ordination number, whilst SIMS imaging shows a low Mn:Si ratio (~1:70) combined with an absence of Mn-rich clusters. Instead, the XANES spectrum agrees very well with that of dilute (isolated) Mn in Si systems, which in turn differs significantly from substitutional Mn:Si [27]. The XAS analysis therefore confirm our interpretation of interstitial doping.

We deduce that once the Mn atoms are released from the organic macrocycle they reach the Si substrate and are introduced into its bulk via various photo-assisted mechanisms. Under equilibrium conditions, incorporation of interstitial Mn into Si typically requires ~2.5-3.2 eV [24], which is amply provided by the 172 nm (7.2 eV) lamps. We note that a thin $SiO_2$ layer re-growth would not block Mn diffusion [28]. Presumably, the macroscopic Mn inhomogeneity arises from surface or bulk inhomogeneities in the original wafer, associated with recombination-enhanced diffusion due to UV-generated electron-hole pairs during illumination [29]. If so, the effects might be diminished by decreasing the UV exposure time and reducing the precursor film thickness.

A major advantage of our technique is its potential for lateral control of dopant positions at the nano-scale. Different MPc packing arrangements on substrates would lead to linear, nearly square or random molecular templates [30, 31]. Fine-tuning of dimensions could be achieved by ligand choice, whilst co-depositing different species could extend the scheme to multiply-doped systems. Accurate control over the time of UV treatment would be essential for the creation of arrays: sufficient to incorporate the Mn atoms but short enough so as not to promote lateral diffusion when illuminated. We note that temperature can also be used to control dopant diffusion. Interstitial Mn diffuses relatively fast at room temperature, quickly disturbing any ordering: Mn would diffuse by a distance of $d \sim (Dt)^{1/2}$ = 1 nm after a time $t$ of 12 seconds (using our derived value of $D$). However, the low thermal budget means such difficulties could be overcome by sample cooling during and after preparation; the use of dry ice (193 K) would reduce Mn diffusion to ~1 nm in a year, whilst even fast diffusing species such as Cu would take ~200 years to migrate 1 nm at 77 K [32].

In conclusion, we have described a versatile, low temperature method to introduce metal dopants into Si. We have shown that a UV source can be used to photo-degrade metal phthalocyanine thin films, and through SIMS and XANES measurements we have demonstrated UV-induced introduction of interstitial Mn into Si. We envisage that this process could be readily extended to other metal/organic/semiconductor combinations. Our procedure is clean, cheap and commercially viable, and so has a large potential for use in the development of nano-electronic devices.

**Acknowledgements**

Financial support from the Research Council UK and the Engineering and Physical Sciences Research Council Basic Technology grant (GR/S23506) is gratefully acknowledged. We thank the Royal Society for a Dorothy Hodgkin research fellowship and a Wolfson Research Merit Award. We acknowledge the ESRF for provision of their facilities and P. Glätzel and T.-C. Weng for assistance. We thank Professors D. van der Marel and C. Renner for providing the MnSi reference sample.

**References**

[1] H. G. Grimmeiss, E. Janzen, and B. Skarstam, J. Appl. Phys. 51, 3740 (1980).

[2] M. L. Reed, N. A. El-Masry, H. H. Stadelmaier, M. K. Ritums, M. J. Reed, C. A. Parker, J. C. Roberts, and S. M. Bedair, Appl. Phys. Lett. 79, 3473 (2001).

[3] H. Ohno, A. Shen, F. Matsukura, A. Oiwa, A. Endo, S. Katsumoto, and Y. Iye, Appl. Phys. Lett. 69, 363 (1996).

[4] D. N. Jamieson, C. Yang, T. Hopf, S. M. Hearne, C. I. Pakes, S. Prawer, M. Mitic, E. Gauja, S. E. Andresen, F. E. Hudson, et al., Appl. Phys. Lett 86, 202101 (2005).

[5] X. L. Yang, Z. T. Chen, C. D. Wang, S. Huang, H. Fang, G. Y. Zhang, D. L. Chen, and W. S. Yan, J. Phys. D:Appl. Phys. 41, 125002 (2008).

[6] S. R. Schofield, N. J. Curson, M. Y. Simmons, F. J. Rue, T. Hallam, L. Obereck, and R. G. Clark, Phys. Rev. Lett. 91, 136104 (2003).

[7] D. Kitchen, A. Richardella, J.-M. Tang, M. E. Flatt, and A. Yazdani, Nature 442, 436 (2006).

[8] J. C. Ho, R. Yerushalmi, Z. A. Jacobson, Z. Fan, R. L. Alley, and A. Javey, Nature Mater. 7, 62 (2008).

[9] I. W. Boyd and J. Y. Zhang, Nucl. Instr. Meth. Phys. Res. B 121, 349 (1997).

[10] N. B. McKeown, Phthalocyanine Materials: Synthesis, Structure and Function (Cambridge University Press, Cambridge, 1998).

[11] M. Bolduc, C. Awo-Affouda, A. Stollenwerk, M. B. Huang, F. G. Ramos, G. Agnello, and V. P. LaBella, Phys. Rev. B 71, 033302 (2005).

[12] H. Wu, P. Kratzer and Matthias Scheffler, Phys. Rev. Lett. 98, 117202 (2007).

[13] S. Heutz, C. Mitra, W. Wu, A. J. Fisher, A. Kerridge, A. M. Stoneham, A. H. Harker, J. Gardener, H.-H. Tseng, T. S. Jones, et al., Adv. Mater. 19, 3618 (2007).

[14] J. H. Sharp and M. Abkowitz, J. Phys. Chem. 77, 477 (1973).

[15] K. R. Rajesh and C. S. Menon, Mater. Lett. 51, 266 (2001).

[16] N. Otha and M. Gomi, J. Appl. Phys. 39, 4195 (2000).

[17] S. Heutz, S. M. Bayliss, R. L. Middletone, G. Rumbles, and T. S. Jones, J. Phys. Chem. B 104, 7124 (2000).

[18] Q.-R. Li, C.-Z. Gu, Y. Di, J. Yin, and J.-Y. Zhang, J. Hazard. Mater. 133, 68 (2006).

[19] A. A. Dakhel, Thin Solid Films 496, 353 (2006).

[20] M. H. Yang, G. Mount, and I. Mowat, J. Vac. Sci. Technol. B 24, 428 (2006).

[21] S. W. Novak and R. G. Wilson, J. Appl. Phys. 69, 463 (1991).

[22] M. Ashida, N. Uyeda, and E. Suito, Bull. Chem. Soc. Jpn. 39, 2616 (1966).

[23] H. Nakashima and K. Hashimoto, J. Appl. Phys. 69, 1440 (1991).

[24] G. M. Dalpian, A. J. R. da Silva, and A. Fazzio, Phys. Rev. B 68, 113310 (2003). 9

[25] H. Chen, W. Zhu, E. Kaxiras, and Z. Zhang, Phys. Rev. B 79, 235202 (2009).

[26] F. H. B. Lima, M. L. Calegaro, and E. A. Ticianelli, J. Electroanal. Chem. 590, 152 (2006).

[27] A. Wolska, K. Lawniczak-Jablonska, M. Klepka, M. S. Walczak, and A. Misiuk, Phys. Rev. B 75, 113201 (2007).

[28] L. Zhang and D. G. Ivey, J. Mater. Res. 6, 1518 (1991).

[29] N. Itoh and A. M. Stoneham, Materials Modification by Electronic Excitation (Cambridge University Press, Cambridge, 2001).

[30] J. Gardener, J. H. G. Owen, K. Miki, and S. Heutz, Surf. Sci. 602, 843 (2008).

[31] M. D. Upward, P. H. Beton, and P. Moriarty, Surf. Sci. 441, 21 (1999).

[32] A. Zamouche, T. Heiser, and A. Mesli, Appl. Phys. Lett. 66, 631633 (1995).

**Figure Captions**

**Figure 1** UV-Vis spectra of a 5 nm thick MnPc fillm on glass (as indicated) and after 172 nm UV exposure for 30 seconds, 1, 2, 5, 10, 20 and 40 minutes. Spectra of clean glass substrates subjected to the same UV dosage have subtracted from the raw data to correct for UV-induced changes in each case. The intensity of the MnPc absorption peaks are seen to decrease with increasing UV dosage, demonstrating degradation of the molecular film. The structure of MnPc is shown in the inset.

**Figure 2** SIMS depth profiles of (a) a 5 nm MnPc film on Si that had been UV irradiated and Pirhana cleaned, then stored in ultra-high vacuum at room temperature for 25 h prior to measurement and (b) the initial silicon wafer. A significant quantity of Mn is observed in the UV exposed sample, which is not present in the original Si wafer.

**Figure 3** (a) Normalised XANES profiles and (b) EXAFS oscillations for MnPc, Mn, Mn silicide and a MnPc film after UV exposure. Dashed and dotted lines show the edge position for $Mn^0$ and MnPc respectively, while the shaded box highlights the region of Mn oxides. The spectra for the implanted sample are noisier due to the smaller concentration of Mn present. The closest agreement is observed between the UV treated sample and the Mn silicide.

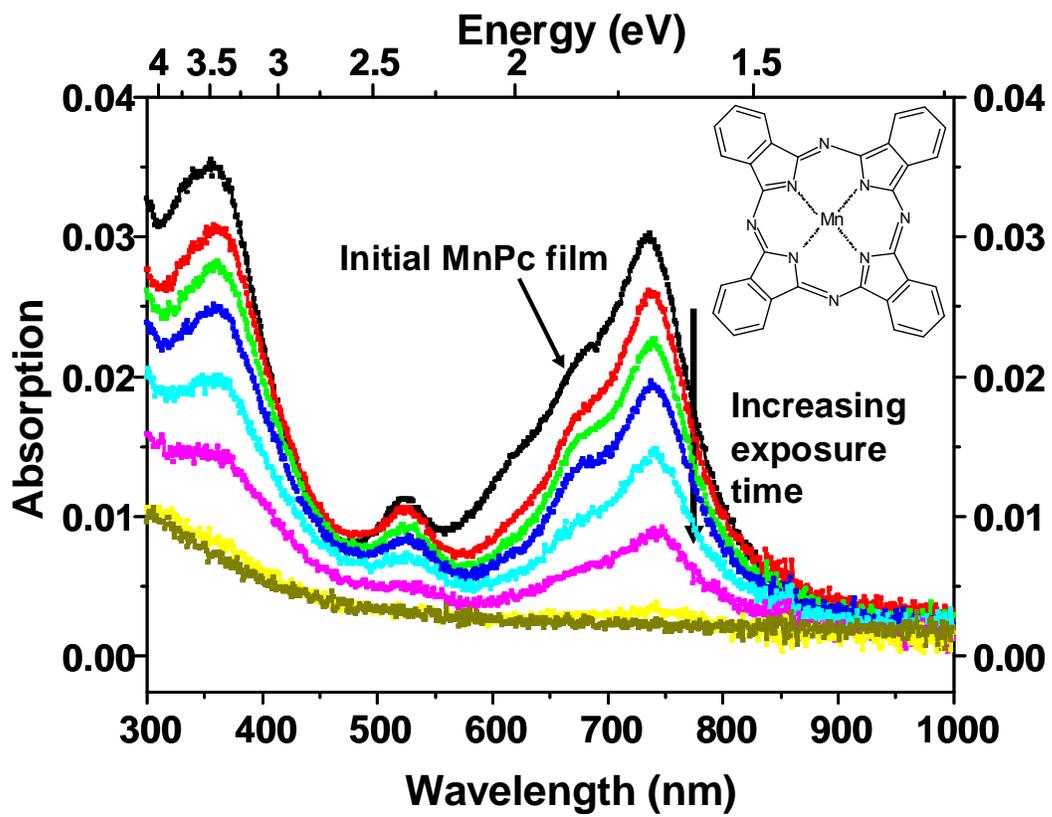

Figure 1

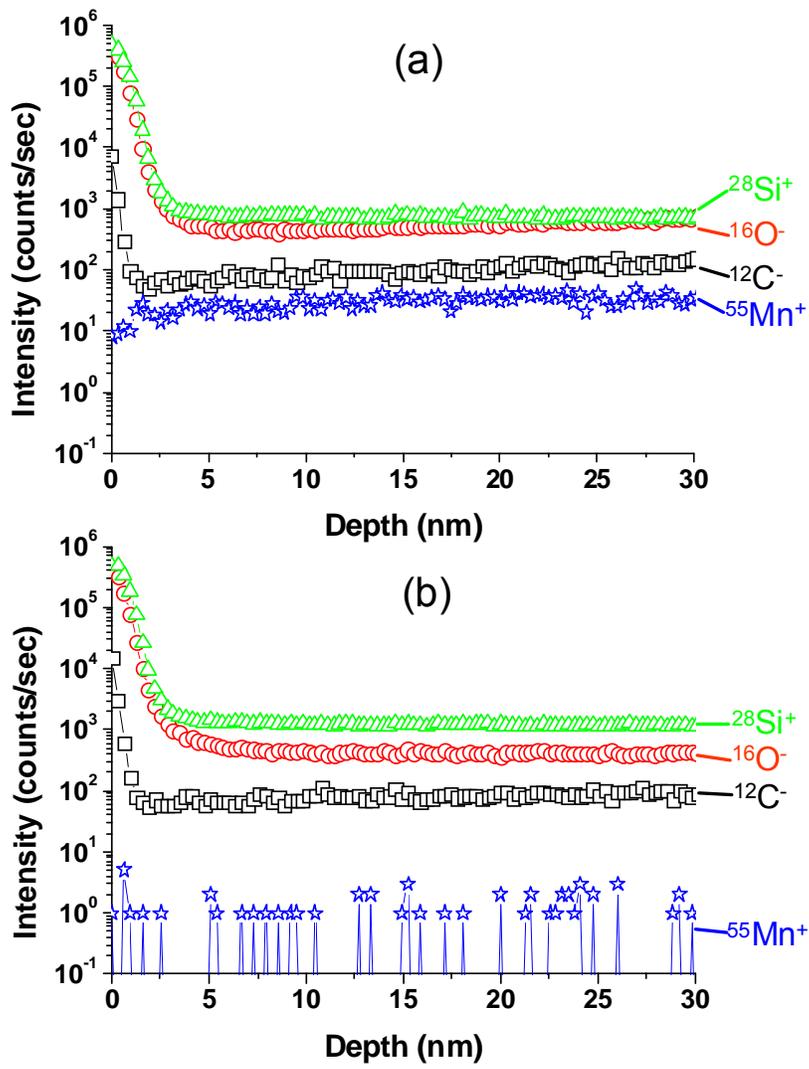

Figure 2

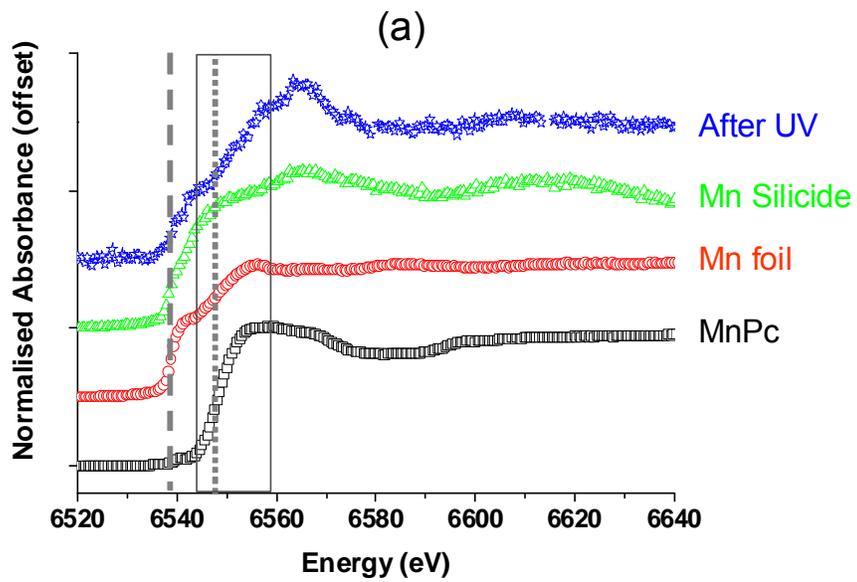
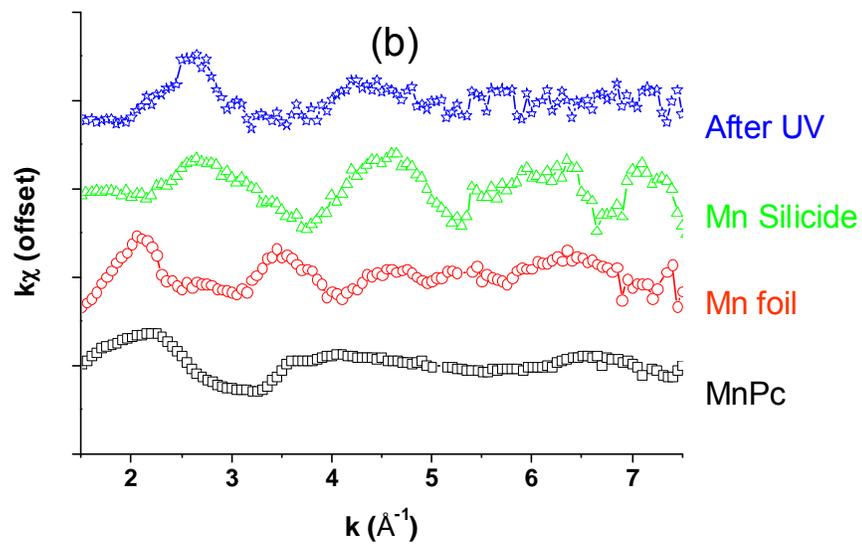

Figure 3